\documentclass[manuscript]{aastex}

\slugcomment{}

\shorttitle{}
\shortauthors{}

\begin{document}

\title{The dusty Nebula surrounding HR~Car: a Spitzer view}


\author{G. Umana, C. S. Buemi and C. Trigilio } \affil{INAF-Osservatorio
Astrofisico di Catania, Via S. Sofia 75, 95123 Catania, Italy }

\author{J. L.  Hora and G. G. Fazio} 
\affil{Harvard-Smithsonian Center for Astrophysics, 60 Garden St. MS-65,
Cambridge, MA 02138-1516, USA}

\and

\author{P. Leto}
\affil{INAF- Istituto di Radioastronomia, C.P.  141, Noto, Italy }
\email{}



\begin{abstract}

We present mid-IR observations of the Galactic 
Luminous Blue Variable (LBV) HR~Car and its
associated nebula carried out with the Spitzer Space Telescope using
both IRAC and IRS, as part of a GTO program aimed to study stellar ejecta
from evolved stars.  Our observations reveal a rich mid-IR spectrum of
the inner nebula showing both solid state and atomic gas signatures.
Strong low-excitation atomic fine structure lines such as $ 26.0 \mu$m
[\ion{Fe}{2}] and $ 34.8 \mu$m [\ion{Si}{2}], indicate, for the first
time, the presence of a PDR in this object class. While the physics
and chemistry of the low-excitation gas appears to be dominated by
photodissociation, a possible contribution due to shocks  can be inferred from the
evidence of gas phase Fe abundance enhancement.

 The presence of amorphous silicates, inferred from the observed
characteristic broad feature at $10 \mu$m located in the inner nebula,
suggests that dust has formed during the LBV outburst.  This is in contrast
with the detection of crystalline dust in other probably more evolved
Galactic LBVs, which is similar to the crystalline dust observed in red
supergiants. This has been considered to be evidence of dust production
during evolutionary phases prior to the outburst.

\end{abstract}





\section{Introduction}

Luminous Blue Variables (LBVs) are luminous, massive stars, which
represent a crucial and relatively short phase of massive star evolution
between core-hydrogen burning O-type stars and helium burning Wolf-Rayet
(W-R) stars. LBVs are quite rare objects in our Galaxy -- a recent census
reported 12 confirmed members and 23 candidates \citep{clark05}.  From an
observational point of view, LBVs are characterized by strong spectral and
photometric variability \citep{vanGenderen01}.  Such variability may
indicate an increase of the mass-loss rate \citep{davidson87} or a change
in the radius of the star \citep{leitherer89}.  Giant eruptions, during
which a large amount of material is released ($10^{-5}-10^{-4}
M_{\odot}yr^{-1}$), have been observed between prolonged periods of
quiescence. Such events have been witnessed very rarely but the presence
of extended circumstellar nebulae (LBVN) around Galactic and extragalactic
LBVs suggests that they are a common aspect of LBV behavior
\citep{nota92}. Most of well known LBVNs contain large quantities of dust:
they have been detected by IRAS, ISO, MSX and ground-based near IR imaging
as having excess IR and spatially resolved shells images
\citep{Tram98,egan_02,clark_03}.

According to recent evolutionary models, a massive star has to lose a
huge quantity of its Main Sequence (MS) mass during its post-MS evolution
until it reaches its W-R phase during which it completely sheds its H
envelope. \cite{lamers01} point out that LBVs and related transition
objects (Ofpe/WN9 stars), which undergo massive mass-loss and eruptive
events, may have a key role in the proposed scenario.  There are, however,
questions related to LBV phenomenon still under debate and, in particular,
the nebula formation mechanism; i.e., the intensity, duration and geometry
of single mass-loss event is not yet established. The full
characterization of mass-loss properties during the LBV phase appears,
therefore, crucial to assess their role in massive star evolution.

Much information on the LBV phenomenon can be derived by the study of the
LBVNs, in particular the detailed analysis of gas and dust content in the
LBVNs can be be used to tackle the problem of their origin.  By studying
the dust composition, it is possible to observe differences between LBVs
that result from differences in the dust forming processes and/or the
nebular abundances. The latter, as the dust chemistry would largely
reflect the chemical composition of the gas phase, would help in
localizing at which phase of stellar evolution the dust has been produced.
Moreover, determining the distribution and the morphology of the gas and
dust in the nebula may provide clues about the mass-loss history
during and/or after the dust formation process.

In order to begin to answer some of these questions, we have started a
systematic study of a sample of LBVNs associated with Galactic LBVs using
the Spitzer Space Telescope.  The study is aimed to detect and resolve the
faint dust shells that have been ejected from the central stars with
mid-IR imaging, and to characterize the mineral composition of LBV ejecta
via mid-IR spectra.  In this paper we present results on the well-known
Galactic LBV HR~Car.


\subsection{HR Car}
HR Car is one of coolest Galactic LBV, classified as B2I.
From a spectral analysis of the Balmer lines \cite{Manchado2003} have recently derived a stellar temperature of $10,000  K$ and a luminosity of 
$5 \times 10^{5} L_{\odot}$, putting HR~Car close to the  $40 M_{\odot }$  evolutionary track in the H-R diagram. 
The distance to HR~Car ($ 5 \pm 1 kpc$) has been derived by \cite{vanGenderen91}, using the reddening-distance method, which is consistent with 
the value obtained from the kinematic measurements of the Carina spiral arm ($5.4 \pm 0.4 kpc$) by \cite{hut91}.

HR~Car is surrounded by a faint, low-excitation nebula which is very
difficult to observe in the optical due to the strong luminosity of the
central object. Despite such difficulties, HR~Car and its associated
nebula have been extensively studied. Among the most recent results,
\citet{weis97} and \citet{not97}, both based on optical high resolution
coronographic imaging and spectroscopy, found that the nebula has a
large-scale bipolar morphology, very reminiscent of the $\eta$ Carina
nebula. Such lobes are expanding in the SE-NW direction (PA=135 degree)
and have a diameter of $\sim 18 \arcsec$ \citep{not97}. On a smaller scale
of a few arcsecs from the central object, a more compact, brighter nebula
is evident, whose morphological details, however, could not be derived
from the coronographic optical ($H_{\alpha}$) images.  \cite{not97}
pointed out some degree of asymmetry, especially in the SE region of the
inner nebula. The presence of some material close to the central object
whose distribution is not symmetric was also inferred from the BVRI
polarization measurements reported by \citet{par00}.

Most of the information on the inner nebula can be obtained at longer
wavelengths, where the difficulties due to the contrast between the bright
central object and the faint nebula can be overcome. \cite{voors00} mapped
the dust (broad-band 10 $\mu$m) and the ionized gas (narrow-band
[\ion{Ne}{2}], 12.8 $\mu$m) of the nebula surrounding HR~Car. The dusty
nebula is quite compact, extending a few arcsecs from the central object,
and reveals an asymmetric morphology. The ionized gas has the same extent
as the dust, but a different distribution, implying a partial screening of
the stellar ionization flux by the circumstellar dust.  The most striking
result is the difference between the morphology observed at smaller
angular scale to that observed in the optical; this may imply that we
are tracing different mass-loss episodes, which probably occur with a
time-dependent geometry.  The same conclusion was reached by
\cite{White00}, who observed HR~Car with the ATCA. The radio images reveal
an extended ionized region whose morphology closely resembles that of the
$H_{\alpha}$ bipolar nebula plus an inner, compact nebula a few arcsecs
across.  The inner nebula is strongly asymmetric and in complete
disagreement with the large scale structure. Another puzzling feature of
this LBV is that the central star doesn't appear to produce a sufficient
quantity of ionizing photons to keep the nebula ionized, even assuming a
previous period of higher temperature for the star.  A possible
explanation that takes in account both the observed degree of
ionization and  the strong asymmetry of the inner nebula is a symbiotic
model, as put forward by \cite{White00}, but the presence of a companion
with the required properties has not yet been confirmed.

\section{Observations and Data Reduction}

The observations presented in this paper were carried out by using the
Infrared Array Camera \citep[IRAC;][]{fazio04} and the Infrared
Spectrograph \citep[IRS;][]{houck04} on board the Spitzer Space Telescope
\citep{werner04}. These observations were part of the GTO program (ID
30188) and were performed on 2006 July 10 (IRAC; AOR 17335040) and 2006
June 28 (IRS; AOR 17335296). Both imaging and spectral observations were
obtained at the central source position ($RA_{2000}=10h22m53.84s,
DEC_{2000}=-59d37m28.4s$), which is determined, with good accuracy, from
previous IR and radio imaging.

\subsection{IRAC 3.6, 4.5, 5.8 and 8.0 $\mu$m Images}

HR~Car was observed with 8 IRAC frames per band using a medium cycling
dither pattern at a frame time of 12 seconds, providing a field of view of
approximately 6x6 arcmin in the final mosaics. The Basic Calibrated Data
(BCD) pipeline version S14.0.0 were used in the reduction, and IRACproc
\citep{schu06} was used to create the IRAC final images at a pixel scale
of 0.6 arcsec/pixel (the original IRAC scale is approximately 1.2
arcsec/pixel).  The source structure revealed by previous mid-IR imaging
\citep{voors00,trams97} is too compact to be well-resolved by IRAC.  
However, in these observations we are sensitive to fainter emission that
is extended further from the star.

\begin{figure}
\epsscale{0.8}
\includegraphics[angle=0,scale=0.8]{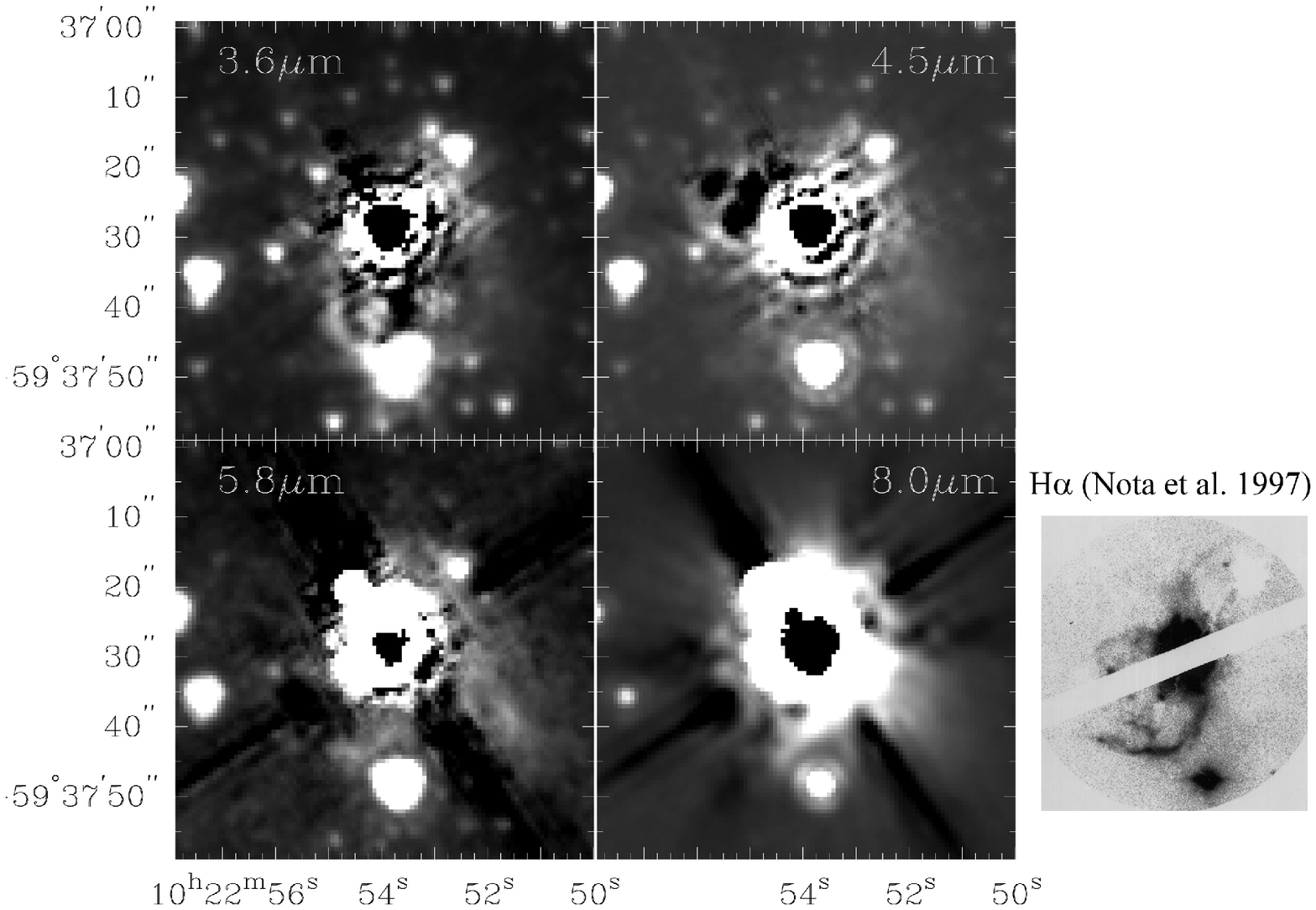}
\caption{The PSF-subtracted images of HR~Car based on the 12 sec frames
obtained in each of the four IRAC channels. The core of the source was
saturated and therefore appears as a black region in the center of the
image. The images were rotated approximately 123.3 degrees counterclockwise
to align them to R.A. and Dec. shown, therefore the array banding,
muxbleed, and bandwidth artifacts \citep{hora04} are also rotated from
their nominal BCD orientation and appear as dark or bright residuals that
are roughly diagonal in the frame. 
The H$\alpha$ image from \citet{not97} is shown to the right of the IRAC
images. There is 
low-level extended emission is visible in the images which is similar to
that seen in the H$\alpha$ image. The residual
rings near the core result from a mismatch between the PSF and
HR~Car images, possibly indicating that the core is slightly
broader than a point source. \label{fig-1}}
\end{figure}

At the 12sec frame time, the central star is saturated in all bands in the
BCD.  The PSF-subtraction technique in IRACproc was used to measure the
flux in each of the bands. The PSF-subtracted mosaics are
shown in Figure \ref{fig-1}, and the flux and magnitudes of the central
star are given in Table \ref{tbl-1}.  The PSF-subtracted images have
residuals, especially in the central regions, which are in part related to
the slightly extended emission around the central star in the mid-IR.  
The photometry in Table \ref{tbl-1} is the flux from the central star only
and not the extended emission which is difficult to estimate close to the
core.  However, one can see evidence in the images for emission on the
same spatial scale as seen in the H$\alpha$ image \citep{not97}, which is
brighter at longer IRAC wavelengths.
 
\subsection{IRS Spectra}

\begin{figure}
\epsscale{1.0}
\plotone{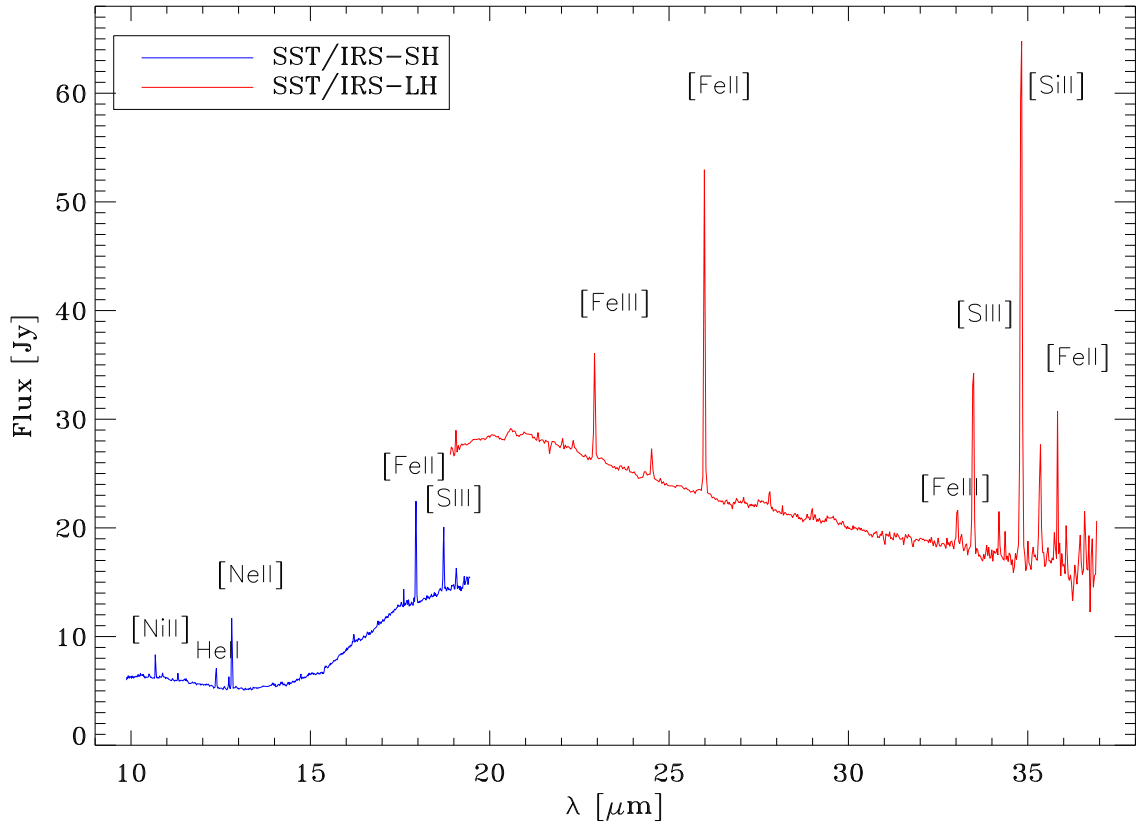}
\caption{IRS High-resolution spectrum of HR~Car. See the electronic edition of the Journal for a color version of this figure.\label{fig2}
}
\end{figure}

The spectral observations were obtained by using the Short-High (SH) and
Long-High (LH) modules ($R\sim 600$), which cover two spectral ranges
going from 9.9 to 19.6 $\mu$m (SH) and from 18.7 to 37.2 $\mu$m (LH). For
each module, three cycles were performed for each nod position in order
to provide rejection of cosmic rays and other transient.  From the 2D
Basic Calibrated Data (IRS pipeline version 15.3.0) three spectra, for
each nod position, were extracted and wavelength and flux calibrated using
the Spitzer IRS Custom Extractor (SPICE) with the extended source
extraction option. Using the Spitzer contributed software SMART
\citep{higdon04}, we cleaned the resulting 1D spectra for residual bad
pixels, spurious jumps, and glitches, and smoothed and merged them into
one final spectrum per module. The resulting high resolution spectra for
each module are shown in Fig.~\ref{fig2}.

\begin{figure}
\epsscale{1.0}
\plotone{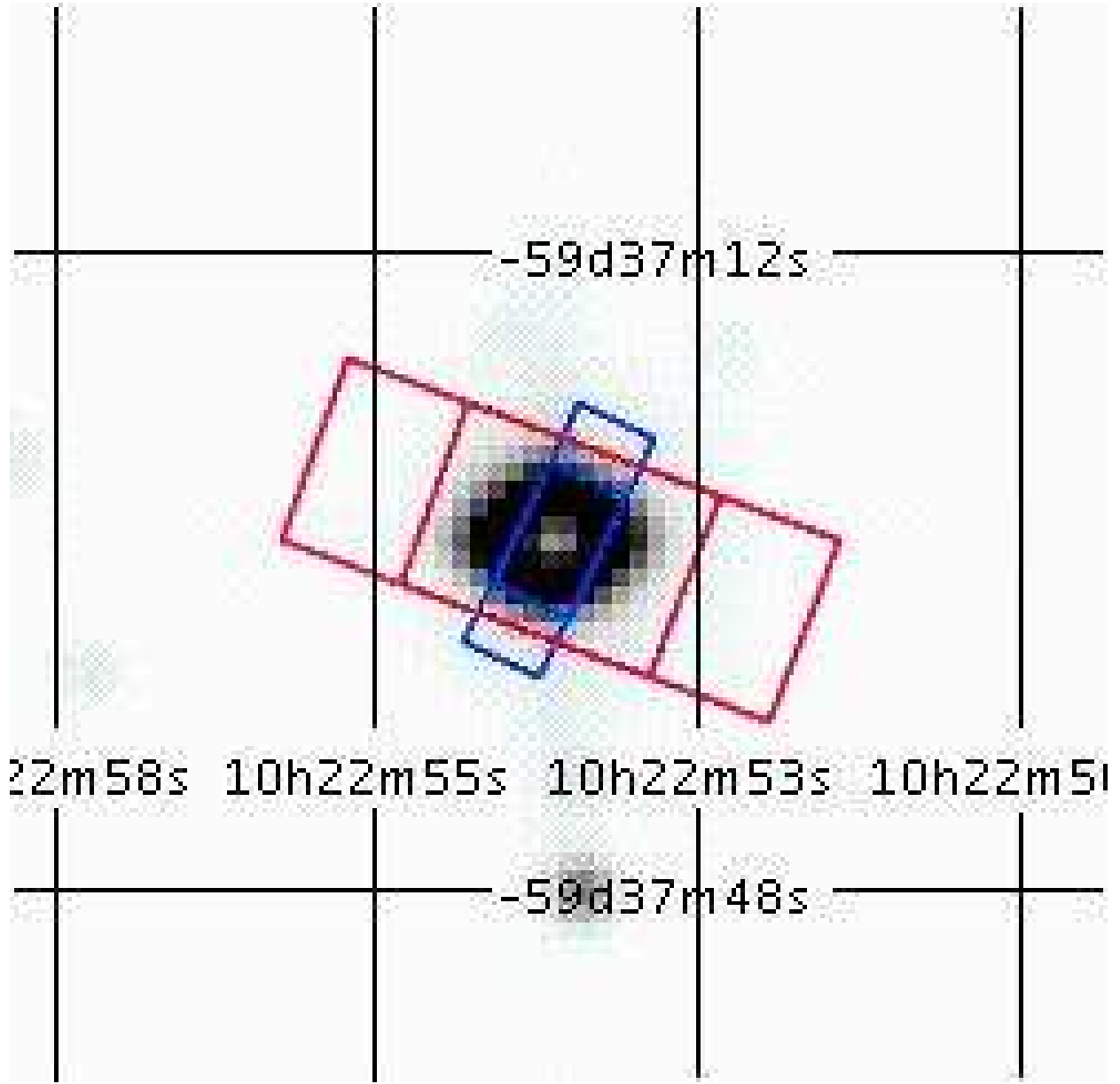}
\caption{Overplot of the IRS spectral slits, for both high resolution modules, on the two Micron All Sky Survey (2MASS) $K_{s}$ image. The smaller aperture (blue box in the electronic edition) corresponds to the SH aperture, the larger one (red box in the electronic edition) is the LH aperture. See the electronic edition of the Journal for a color version of this figure.\label{fig1}
}
\end{figure}

The infrared emission from HR~Car consists of narrow emission lines and a
solid state feature, at about 10 $\mu$m, superimposed on a thermal
continuum.  There is an expected mismatch between the SH and the LH
spectra which is usually attributed to the difference in the two module
apertures, as the SH has a slit size of $4.7 \times 11.3 \arcsec$ and the
LH has a slit size of $11.1 \times 22.3 \arcsec$. The overplot of the IRS
spectral slits for each high resolution module on the two Micron All Sky
Survey (2MASS) $K_{s}$ (Fig.~\ref{fig1}) reveals that the central object
is well contained in both apertures while only part of the inner nebula is
contained in the SH slit. To quantify the correction for aperture losses,
we can take advantage of the small spectral overlap between the two
modules (from 18.7 to 19.6 $\mu$m) deriving a scaling factor of 1.87, at
that wavelength range, to align the SH spectrum to the LH spectrum.

However, we cannot a priori rule out a wavelength dependence of the
missing flux, and a simple scale factor may be an oversimplification.  We
have therefore retrieved from the ISO Data Archive (IDA v9.0) a SWS
spectrum (TDT=24900321), obtained over the full SWS grating range
($2.38-45.2 \mu$m), with a spectral resolution comparable to that of the
IRS observations ($\sim 600$), and with instrumental apertures ranging
from $14\arcsec \times 20\arcsec$ ($2.4-12 \mu$m) to $20\arcsec \times
33\arcsec$ ($29-45 \mu$m). Therefore, in the SWS observations the whole
inner nebula was included in the aperture. This spectrum was already
presented by \cite{lam96}, but the data we retrieved from the archive are
the result of further processing beyond the original pipeline with new,
more refined algorithms (Highly Processed Data Products), providing a much
improved spectrum with respect to the published one \citep{sloan03}.

\begin{figure}
\epsscale{1.0}
\plotone{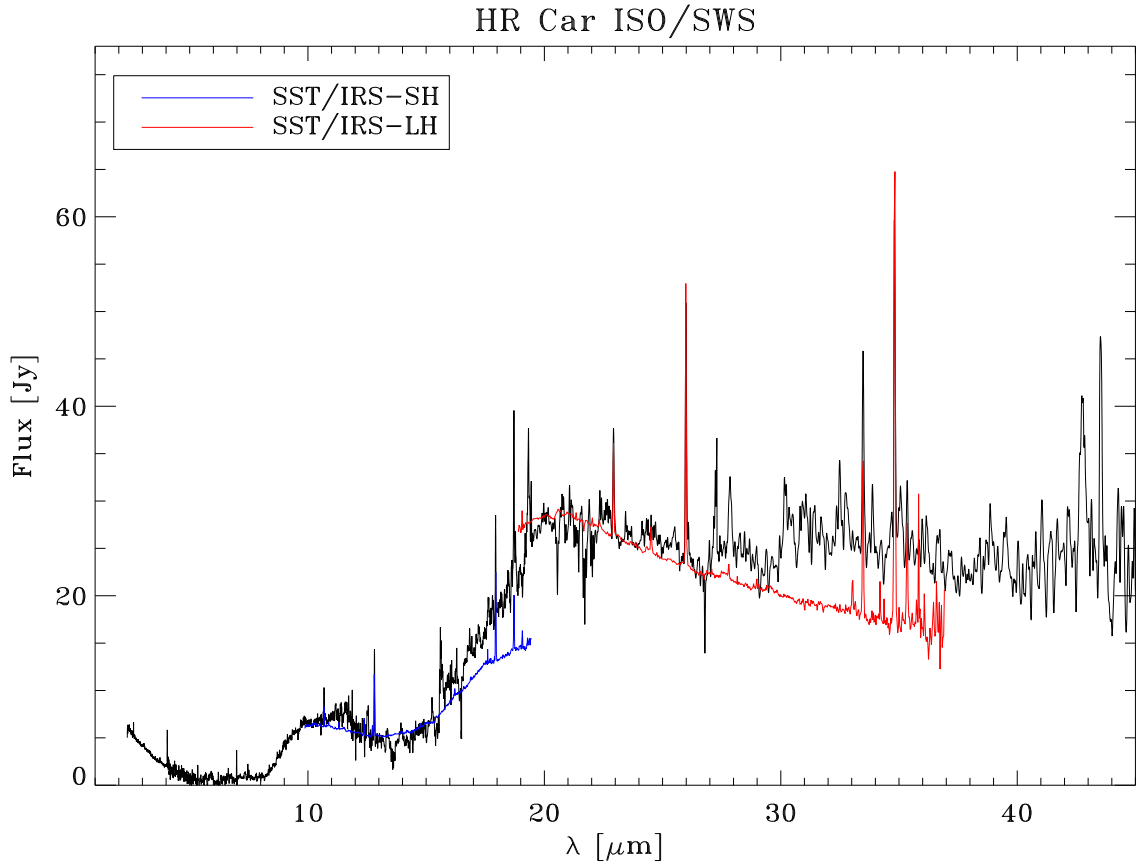}
\caption{The IRS/SH and LH spectra superimposed to the ISO-SWS01 archive spectrum.
See the electronic edition of the Journal for a color version of this figure. \label{fig4}}
\end{figure}

In Fig. \ref{fig4} the two Spitzer-IRS spectra are superimposed on the ISO
data. A good agreement is evident up to $\sim 10 \mu$m. After that, there
is a mismatch, with the amplitude increasing with wavelength, indicating
that a simple scale factor cannot account for aperture losses. The
IRS-LH is again in good agreement with the ISO spectrum up to $\sim 27
\mu$m. At higher wavelengths in the ISO spectrum, there are additional
apparent features which have been reported by \cite{lam96}, but these
features are completely absent in the IRS spectrum. We will further
discuss the implications of these observed differences between the spectra
in the following section.

\section{Solid state features}

The mid-IR spectrum of HR~Car reveals a prominent feature that peaks at
around $10 \mu$m that is probably related to the $ 9.7 \mu$m feature
due to the Si-O stretching mode in amorphous silicates. This feature was
also noted by \cite{lam96}.  The presence of amorphous silicate in the
stellar ejecta indicates an oxygen-rich ($C/O \leq 1$) environment, as we
expect for LBV ejecta being enriched by CNO processed material
\citep{not97}.

In the comparison of the IRS spectrum with the ISO SWS spectrum (Fig.\ref{fig4}),
we notice a reasonably good match between the two spectra at around $10
\mu$m. Since the two instruments have different apertures, the fact that
in both spectra we see the same features with the same intensity implies
that amorphous silicates must form very close to the central object, well
inside the inner nebula, in a region whose extension must be comparable to
or less than the IRS-SH aperture.  At longer wavelenghts, we find that
in the ISO spectrum there is a strong contribution beyond 27 $\mu$m that
is completely absent in the IRS spectrum.  However the ISO spectrum is
quite noisy in that wavelength range and it is difficult to assess the
reliability of the observed broad structures. If we assume that the
observed difference is related to the aperture, we may conclude that the
emission features arise from a region located outside of the inner nebula,
otherwise it should have been detected in the IRS spectrum. The features
must be present in a region which is, at least partly, inside the area
seen by the SWS aperture. The nature of such a contribution is not clear.
\cite{lam96} and \cite{wat97} indicate a solid state origin, but no firm
conclusion on the possible carrier/s was reached.

At this time, we can only speculate on the possible location of this
material, but we cannot discriminate whether the spectral features are
connected to some outer region of the HR~Car nebula or to the ISM
surrounding it.

\section{The emission line spectrum}

The mid-IR spectrum of HR~Car shows many fine structure lines that, once
the mechanisms responsible of the observed emission have been identified,
may provide important constraints on the physical conditions in the line
emitting region.  The line-fitting routine in SMART was used to identify
and measure the narrow lines. For each line a local continuum was defined
by a single-order fit to the baseline and the line was then fitted using a
single Gaussian.

The most prominent lines are labeled in Fig~\ref{fig2}, while the 
identification and the measured fluxes, including weaker 
lines $(S/N \leq 20$) are summarized in
Table~\ref{tbl-2}. Possible multiple identifications are also indicated. 
High ionization potential lines such as those of
[\ion{S}{3}] and [\ion{Fe}{3}] can only originate in the \ion{H}{2} region around
the star, while the fine structure lines such as those of [\ion{Fe}{2}]
and [\ion{Si}{2}] can form in a photodissociation region (PDR) surrounding
the ionized part of the LBVN, where the gas is predominantly neutral but
whose chemistry and thermal balance is controlled by FUV photons, i. e.
with energies between 6 and 13.6 eV \citep{Tielens85}. PDRs are bright in
the far-IR dust continuum, in the PAH emission features, in some infrared
fine-structure lines and in rotational lines of CO. CO emission detected
in other LBV \citep{Nota02, Rizzo08} supports the hypothesis of the
presence of PDRs in the LBVNs.

\subsection{[\ion{S}{3}] lines}

The [\ion{S}{3}]  33.48 $\mu$m and [\ion{S}{3}] 18.71 $\mu$m lines can be
used to derive the electron density in the ionized nebula associated with
HR~Car. Those lines originate from levels close in energy, and therefore
their ratio will have almost no dependence on temperature, being only
sensitive to  their transition probabilities or collisional
de-excitation, which depend on density.

The [\ion{S}{3}] 18.71 $\mu$m line is located in the overlap between the
two high resolution modules. We can therefore safely apply the correction
factor of 1.87 to take the aperture losses into account.  From the
observed ratio of the $ 33.48\mu$m line flux to the $ 18.71 \mu$m line
flux, we derive, assuming a electronic temperature of $T_{e}\sim
10^{4}~K$, an electronic density of the \ion{H}{2} region $n_{e} \geq
10^{3} cm^{-3}$ \citep{houck84}.  Because this line ratio is insensitive
to temperature, the adopted value of $T_{e}$ will not affect the value of
$n_{e}$. The derived value of $n_{e}$ is in good agreement with those
derived by \citet{not97}, who obtained, from [\ion{S}{2}] line ratios, a
gradient in the electron density $n_{e}$ ranging from $\sim 10^{4}
cm^{-3}$ in the center of the nebula to a value of $ \sim 10^{3} cm^{-3}$
at a distance from the core of $\sim 5 \arcsec$. A similar value for the
electron density ($n_{e}=3200 cm^{-3}$) was derived from the central core
of the ionized nebula ($6 \arcsec \times 8 \arcsec$) by White (2000) from
radio measurements.

The agreement of our density determination with those derived by other
authors in other spectral regions also supports the adopted scaling factor
applied to the SH spectrum to account for aperture losses. The
non-corrected $ 18.71 \mu$m line flux would lead to a [\ion{S}{3}] lines
ratio that implies a very low density \ion{H}{2} region ($n_{e}=10-100
cm^{-3}$), which is not consistent with results from visible and radio
observations.

\subsection{[\ion{Fe}{2}] and [\ion{Si}{2}] lines}

The [\ion{Fe}{2}] $ 26.0 \mu$m and [\ion{Si}{2}] $ 34.8 \mu$m lines may
arise from the ionized region of the LBVN (\ion{H}{2} region) and/or from
the associated PDR. Therefore, we cannot a priori exclude a mixed
contribution from these different regions to the observed lines
intensities. The use of these lines as diagnostics of the physical
condition in the PDR is limited by the knowledge of the real PDR
contribution.  In a recent paper, \cite{Kaufman06} modeled the intensities
of typical infrared fine-structure lines that may originate from both the
\ion{H}{2} region and the associated PDR in order to evaluate the contribution to
such diagnostics from the two different regions. They concluded that, in
the hypothesis of thermal pressure balance between the \ion{H}{2} regions
and the PDR and normal metallicity, the fine structure line of
[\ion{C}{2}] at $ 158.0 \mu$m is always dominated by the PDR, while the
[\ion{Fe}{2}] $ 26.0 \mu$m and [\ion{Si}{2}] $ 34.8 \mu$m emission is
dominated by the PDR if the \ion{H}{2} region has a high electron
density $ n_{e} \geq 10 cm^{-3}$. A similar conclusion was reached by
\cite{Abel05}.

Assuming the $ n_{e} $ for the \ion{H}{2} region associated with HR~Car as
derived from the [\ion{S}{3}] line ratio, we conclude that the
[\ion{Fe}{2}] $ 26.0 \mu$m and [\ion{Si}{2}] $ 34.8 \mu$m lines observed
in the mid-IR spectrum of HR~Car are PDR-dominated and can be safely used
to derive the physical conditions of this region. For this purpose we need
to compare the observed line intensities with those predicted by the
theoretical PDR models.
 
The physics and chemistry of a PDR are governed by the incident FUV ($ 6
eV \leq h\nu \leq 13.6 eV$) radiation, often expressed as relative to the
background interstellar radiation field assumed to be $ 1.6 \times 10^{-3}
erg~cm^{-2} s^{-1}$: \begin{equation} G_{o}= \frac{ \int_{912 \AA}^{2067 \AA}
I_{\lambda} d\lambda}{ 1.6 \times 10^{-3} (erg~cm^{-2} s^{-1})}
\end{equation} and by the total density $n_{o}$ ($cm^{-3}$).

\begin{figure}
\plotone{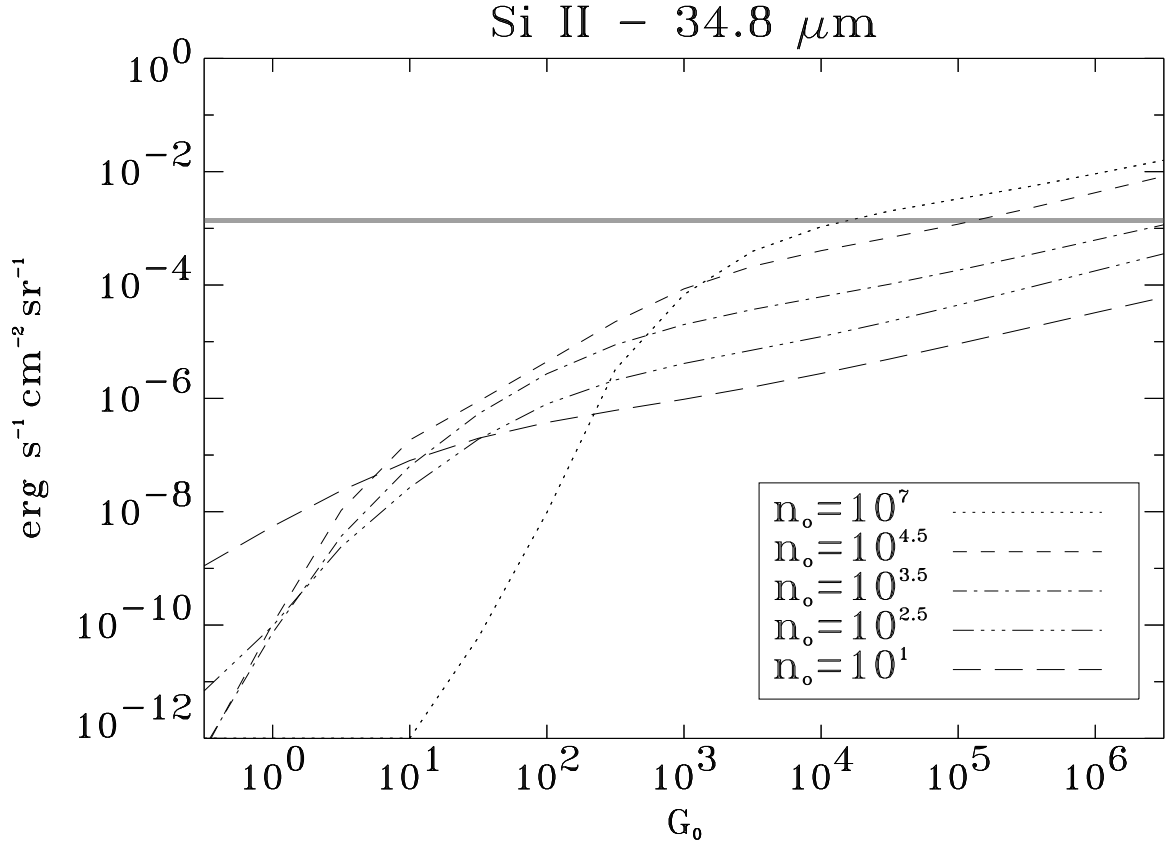}
\caption{The line intensity for
[\ion{Si}{2}] $ 34.8 \mu$m computed with
the model of \cite{Kaufman06} is shown as function of $G_{o}$ for several
values of $n_{o}$.  The shaded region indicates our line measurement with
associated uncertainty.\label{fig-3a}
}
\end{figure}

\begin{figure}
\plotone{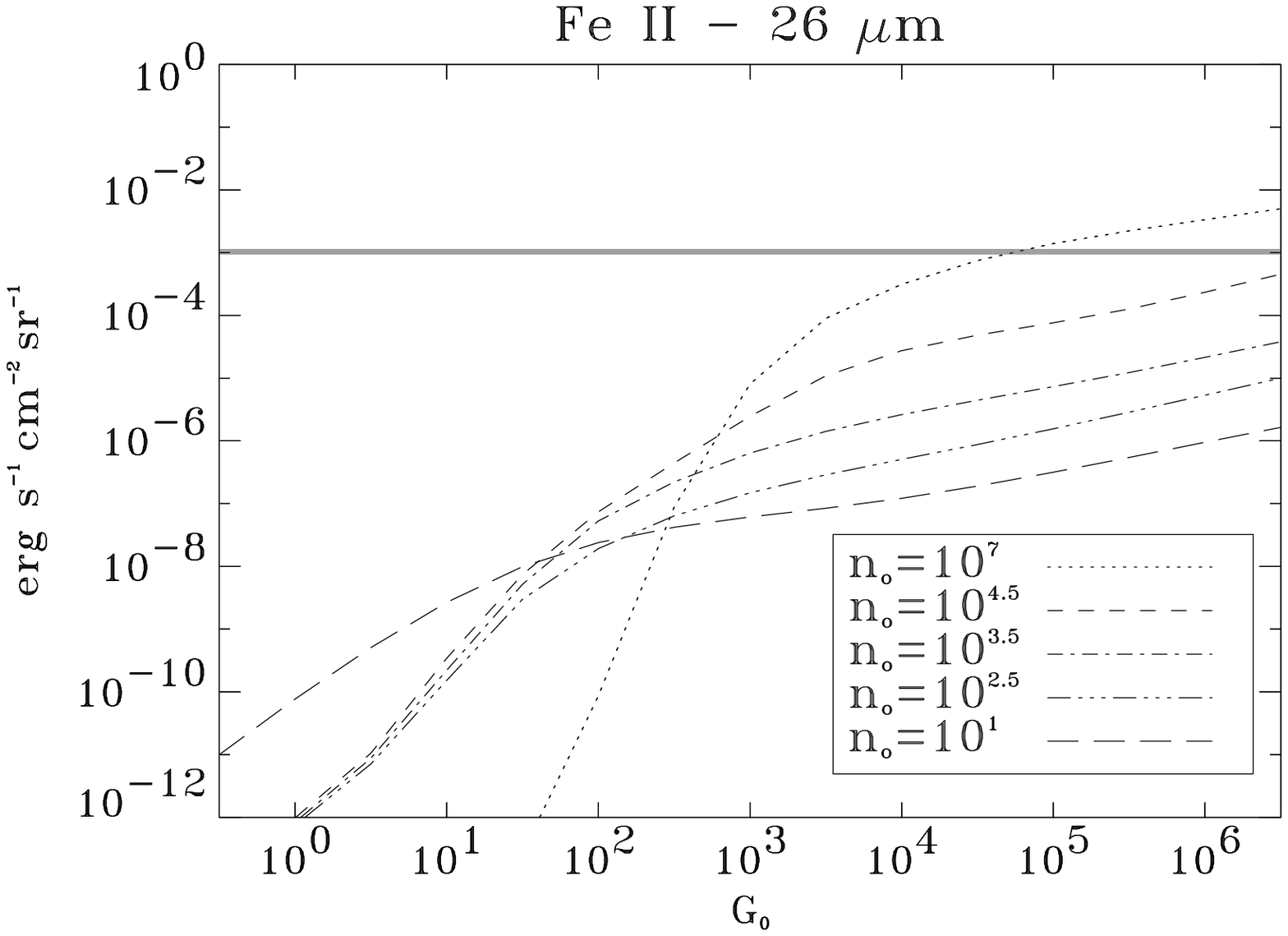}
\caption{Tha same as fig.~ \ref{fig-3a} but for [\ion{Fe}{2}] $ 26.0 \mu$m line.
\label{fig-3b}
}
\end{figure}

In Figures \ref{fig-3a} and \ref{fig-3b}, the line intensities for
[\ion{Si}{2}] $ 34.8 \mu$m and [\ion{Fe}{2}] $ 26.0 \mu$m computed with
the model of \cite{Kaufman06} are shown as function of $G_{o}$ for several
values of $n_{o}$. The measured fluxes of the [\ion{Fe}{2}] $ 26.0 \mu$m
and [\ion{Si}{2}] $ 34.8 \mu$m lines correspond to a line intensity of $
1.04 \times 10^{-3} (erg~cm^{-2} s^{-1} sr^{-1}$)  and $ 1.42 \times
10^{-3} (erg~cm^{-2} s^{-1} sr^{-1}$) respectively, if we assume that the
PDR completely fills the LH aperture ($\Omega_{PDR}=0.57 \times
10^{-8} sr$). The shaded regions in each figure indicate our line
measurement with the associated uncertainty.

Different combinations of the space parameters ($G_{o}$, $n_{o}$) can
explain the observed line intensity of [\ion{Si}{2}] at $ 34.8 \mu$m,
going from a high density ($n_{o}\sim 10^{6}$) and a moderate incident FUV
radiation $log(G_{o})\sim 3.5 $ to a low density ($n_{o}\sim 10^{3.5}$)
and a high incident FUV radiation $log(G_{o})\sim 6 $ (Figure
\ref{fig-3a}). On the other hand, to explain the observed line intensity
of [\ion{Fe}{2}] at $ 26.0 \mu$m, a very high density ($n_{o} \geq
10^{7}$) and a high incident FUV radiation ($log(G_{o})\geq 4.5 $) are
necessary.

Because the [\ion{Fe}{2}] line at $ 26.0 \mu$m and the [\ion{Si}{2}] line
at $ 34.8 \mu$m are optically thin \citep{Tielens85}, the modeled line
intensities scale linearly with the adopted gas phase abundance. This
means that to have physical conditions of the PDR consistent with both the
observed line intensities, the gas phase Fe abundance in the nebula should
be higher than the values adopted by the \cite{Kaufman06} standard model
typical for the diffuse ISM ($1.7 \times 10^{-7}$).

It is quite difficult to quantify the gas phase Fe abundance necessary  to reproduce the observed  $ 26.0 \mu$m [\ion{Fe}{2}] line 
intensity without detailed modelling.
The PDR model of \cite{Tielens85}, which assumes a gas phase Fe abundance of $2.5 \times 10^{-7}$, foresees a
$ 26.0 \mu$m [\ion{Fe}{2}] line intensity of $4 \times 10^{-4} ~erg~cm^{-2} s^{-1} sr^{-1}$, for $G_{o}=10^{5}$ and  $n_{o}=10^{4..5}$.
Therefore, with an increment of $\sim 50 \%$ of the  gas phase Fe abundance, 
the expected $ 26.0 \mu$m [\ion{Fe}{2}] line intensity is enhanced of a factor 
of four, shifting the model curves shown in Fig~\ref{fig-3b} towards our observed values.
\section{Conclusions}

We have presented mid-IR observations of the Galactic LBV HR~Car and its
associated nebula carried out with Spitzer using both IRAC and IRS, as
part of a GTO program aimed to study stellar ejecta from evolved stars.

The central object in HR~Car is too bright in the mid-IR and the warmer
inner dusty nebula too compact to get reliable mapping with IRAC;  still,
the adopted observing strategy has allowed us to observe the presence of
faint structures, brighter at longer IRAC wavelengths, whose extension is
comparable to those of inner ionized nebula traced by $H_{\alpha}$ and the
radio. 
The more extended nebula,  which, in the mid-IR, can be detected only through 
its free-free continuum (no strong recombination lines are evident in 
the ISO spectrum for $\lambda \leq 10 \mu m$),
is expected to be too faint to be observable at the integration time used.

We have further obtained a mid-IR spectrum of the inner nebula and a few
arcsecs across the central object. The spectrum reveals quite a number of
emission lines, some of which are formed in the ionized part of the
nebula.  Particularly interesting is the presence of low-excitation
atomic fine structure lines such as $ 26.0 \mu$m [\ion{Fe}{2}] and $ 34.8
\mu$m [\ion{Si}{2}], lines whose detection may indicate the presence of a
PDR.  The comparison between the observed line intensities and those
derived from recent PDR models supports the existence of a PDR surrounding
the ionized fraction of the nebula. However, to obtain a PDR with physical
characteristics that are able to reproduce both [\ion{Fe}{2}] and
[\ion{Si}{2}] line intensities, a Fe gas phase abundance higher than that
of the ISM is necessary.

In the ISM, Fe is usually highly depleted from the gas phase because of
condensation onto dust grains. In this context, a higher Fe gas phase
abundance with respect to that of the ISM may be considered as indirect
evidence of shocks occurring in the nebula resulting in grain processing
that release Fe atoms into the gas phase.

Such an enhanced Fe abundance has been observed in the radio Arc Bubble of
the Galactic Center and has been interpreted in terms of additional Fe
released to the gas phase as a consequence of grain destruction by shocks
\citep{Simpson07}. Such shocks are probably produced in the strong stellar
winds of the Quintuplet Cluster stars, which also includes a LBV, the
Pistol star. Shock-heating has been proposed by \cite{Smith02} to explain
the bright $ 1.643 \mu$m [\ion{Fe}{2}] line observed in a small sample of
LBVs. While the presence of such a line seems to be a common property of
LBV nebulae, its possible excitation mechanism has not been clearly
established.  We conclude that in the case of HR~Car, the
photodissociation dominates the physics and the chemistry of its outer
envelope, but shocks also play an important  role as inferred from the enhanced gas phase Fe abundance. 

The analysis of the IRS spectrum has allowed us to derive some clues on
the mineralogy and spatial distribution of the dust. There is a strong
signature of amorphous silicate, which appears to be localized within
$\sim 10 \arcsec$ ($0.3 pc$, assuming a distance of 5.4kpc) from the central object, in a region spatially
coincident with the optical, IR and radio inner nebula. Our result is
consistent with the amorphous silicates being localized in the compact
($\sim 3 \arcsec$) nebula mapped at 10 $\mu$m by \cite{voors00}.
Such  a strong emission band is probably also responsible for the diffuse emission detected in the $ 8.0 \pm 1.5 \mu m$ IRAC band.
The spatial coincidence among the IRAC structures with the inner 
$H_{\alpha}$ nebula reported by 
\cite{not97} and \cite{weis97}  strongly supports our conclusion that ionized gas and amorphous silicate are localized within the same area.

Relative to the region of the nebula covered by the IRS apertures, we
cannot confirm the presence of crystalline features indicated by previous
ISO observations. The presence of amorphous silicates, together with the
lack of crystalline silicates in the inner part of the nebula, suggest that
there has been recent dust formation in HR~Car that has condensed during
the LBV eruptions.  This is in contrast with other Galactic LBVs, such as
AR~Car and WR751, where crystalline dust has been observed. The similarity
of the crystalline dust in those objects to the dust observed in red
supergiants has been considered evidence of dust production in a previous
evolutionary phase of these massive stars, prior to the LBV outbursts
\citep{waters98}.
\acknowledgments
We thank the referee for her/his comments and suggestions. 
This work is based in part on observations made with the Spitzer Space
Telescope, which is operated by the Jet Propulsion Laboratory, California
Institute of Technology under NASA contract 1407. Support for this work
was provided by NASA through Contract Number 1256790 issued by
JPL/Caltech. Support for the IRAC instrument was provided by NASA through
Contract Number 960541 issued by JPL.  The IRS was a collaborative venture
between Cornell University and Ball Aerospace Corporation funded by NASA
through the Jet Propulsion Laboratory and Ames Research Center. SMART was
developed at Cornell University and is available through the Spitzer
Science Center at Caltech. This publication makes use of data products from the Two Micron All Sky Survey, which is a joint project of the University of Massachusetts and the Infrared Processing and Analysis Center/California Institute of Technology, funded by the National Aeronautics and Space Administration and the National Science Foundation.

{\it Facilities:} \facility{Spitzer (IRAC)}, \facility{Spitzer (IRS)}.

\clearpage

\clearpage
\clearpage
%

\clearpage


\clearpage


\clearpage

\clearpage

\clearpage

\begin{table}
\begin{center}
\caption{HR~Car Fluxes measured in the IRAC channels\label{tbl-1}}
\begin{tabular}{ccr}
~\\
\tableline\tableline
IRAC Band & Flux Density (Jy)  & Mag\\
\tableline
    3.6   &       0.985  & 6.14\\
    4.5   &       0.848  & 5.82\\
    5.8   &       0.714  & 5.51\\
    8.0   &       0.910  & 4.62\\
\tableline
\end{tabular}
\end{center}
\end{table}

\clearpage

\begin{table}
\begin{center}
\caption{Line Identifications and Measured Fluxes\label{tbl-2}}
\begin{tabular}{crr}
~\\
\tableline\tableline
Identification & $\lambda (\mu m)$  & Flux\tablenotemark{a} \\
\tableline
$\left[ Ni~II \right]$    & 8.68 & 193\tablenotemark{b}  \\
$ HeII/Si IV$  & 10.49 & 29\tablenotemark{c}\\
$ HI$  & 10.85 & 29\tablenotemark{c}\\
$ HeII$  & 11.30 & 12\tablenotemark{c}\\
$ HeII$  & 12.30 & 124\tablenotemark{b}\\
$\left[ Ni~II \right]$    & 12.72 & 50\tablenotemark{c}  \\
$\left[ Ne~II  \right]$   & 12.81 &  374\tablenotemark{b}  \\
$\left[ Co~II  \right]$   & 14.74 &  14\tablenotemark{b}  \\
$ HeII/HI$  & 16.29 & 46\tablenotemark{c}\\
$ HI$  & 17.60 & 8.2\tablenotemark{c}\\
$\left[ Fe~II \right]$   & 17.93 &  268\tablenotemark{b}  \\
$\left[ S~III \right]$    & 18.71 &  303\tablenotemark{b} \tablenotemark{d} \\
$ HI$  & 19.04 & 50\tablenotemark{c}\\
$\left[ Fe~III \right]$   & 22.92 &  255\tablenotemark{b}  \\
$\left[ N~V \right]$   & 24.31 &  18\tablenotemark{c}  \\    
$\left[ Fe~II  \right]$   & 24.51 &  84\tablenotemark{b}  \\
$\left[ Fe~II   \right]$   & 25.98 &  594\tablenotemark{b}  \\
$HI$   & 27.80 &  25\tablenotemark{c}  \\    
$\left[ Fe~III \right]$   & 33.03 &  50\tablenotemark{b}  \\  
$\left[ S~III \right]$    & 33.48 &  280\tablenotemark{b}  \\
$\left[ Si~II \right]$   & 34.81 &  813 \tablenotemark{b} \\
$\left[ Fe~II \right]$   & 35.34 &  157\tablenotemark{b}  \\
\tableline
\end{tabular}
\tablenotetext{a}{Observed fluxes in units of $10^{-14}erg~cm^{-2} s^{-1}$}
\tablenotetext{b}{Flux uncertainties $\leq 10\%$}
\tablenotetext{c}{Flux uncertainties between $10\%$ and  $20\%$}
\tablenotetext{d}{The measured flux has been corrected for the scale factor
1.87}
\end{center}
\end{table}

\end{document}